\documentclass[twocolumn,secnumarabic,amssymb, nobibnotes, aps, prc, superscriptaddress, nobalancelastpage,longbibliography]{revtex4-2}

 \usepackage{graphicx}
\usepackage{bm}% bold math
\usepackage{amsmath} 
\usepackage{braket} 
\usepackage{epsfig}
\usepackage{tensor}
\usepackage{CJKutf8}
\usepackage[version=4]{mhchem}
\usepackage{titlesec}
\usepackage{ifthen}
\usepackage{sidecap}
\usepackage{listings}
\usepackage{array}
\usepackage{floatrow}
\usepackage{enumitem}
\usepackage[para,online,flushleft]{threeparttablex}
\usepackage{float}
\floatstyle{plaintop}
\restylefloat{table}

\usepackage{hyperref}
\hypersetup{breaklinks=true,colorlinks=true,linkcolor=blue,citecolor=blue,filecolor=magenta,urlcolor=cyan}

\usepackage[all]{hypcap}

\newcommand{\nuc}[2]{\hbox{$^{#1}$#2}}

\usepackage{xcolor}
\definecolor{pastelgray}{rgb}{0.81, 0.81, 0.77}
\definecolor{beaublue}{rgb}{0.9, 0.9, 0.93}

\usepackage{orcidlink}
\usepackage{tikz,xcolor,hyperref}

\definecolor{lime}{HTML}{A6CE39}
\DeclareRobustCommand{\orcidicon}{
	\begin{tikzpicture}
	\draw[lime, fill=lime] (0,0) 
	circle [radius=0.16] 
	node[white] {{\fontfamily{qag}\selectfont \tiny ID}};
	\draw[white, fill=white] (-0.0625,0.095) 
	circle [radius=0.007];
	\end{tikzpicture}
	\hspace{-2mm}
}

\foreach \x in {A, ..., Z}{%
	\expandafter\xdef\csname orcid\x\endcsname{\noexpand\href{https://orcid.org/\csname orcidauthor\x\endcsname}{\noexpand\orcidicon}}
}

\makeatletter
\def\@bibdataout@aps{%
\immediate\write\@bibdataout{%
@CONTROL{%
apsrev41Control%
\longbibliography@sw{%
    ,author="08",editor="1",pages="1",title="0",year="1"%
    }{%
    ,author="08",editor="1",pages="1",title="",year="1"%
    }%
  }%
}%
\if@filesw \immediate \write \@auxout {\string \citation {apsrev41Control}}\fi
}
\makeatother

\renewcommand{\vec}[1]{\mbox{\boldmath $#1$}}

\newcolumntype{Y}{>{\centering\arraybackslash}X}

\begin{document}
\begin{CJK*}{UTF8}{gbsn}

\title{Puzzling $B(E2;0^+\rightarrow 2^+)$ strength in the proton dripline nucleus $^{36}$Ca}

\author{Z. C. Xu (许志成)\,\orcidlink{0000-0001-5418-2717}}\email{Email: xuzhicheng@fudan.edu.cn}
\affiliation{Key Laboratory of Nuclear Physics and Ion-beam Application (MOE), Institute of Modern Physics, Fudan University, Shanghai 200433, China}
\affiliation{Shanghai Research Center for Theoretical Nuclear Physics,
NSFC and Fudan University, Shanghai 200438, China}

\author{S. M. Wang (王思敏)\,\orcidlink{0000-0002-8902-6842}}\email{Email: wangsimin@fudan.edu.cn}
\affiliation{Key Laboratory of Nuclear Physics and Ion-beam Application (MOE), Institute of Modern Physics, Fudan University, Shanghai 200433, China}
\affiliation{Shanghai Research Center for Theoretical Nuclear Physics,
NSFC and Fudan University, Shanghai 200438, China}

\author{T. Beck\,\orcidlink{0000-0002-5395-9421}}
\altaffiliation[Present address: ]{LPC Caen, 14050 Caen Cedex, France}
\affiliation{Facility for Rare Isotope Beams, Michigan State University, East Lansing, Michigan 48824, USA}

\author{A. Gade\,\orcidlink{0000-0001-8825-0976}}
\affiliation{Facility for Rare Isotope Beams, Michigan State University, East Lansing, Michigan 48824, USA}
\affiliation{Department of Physics and Astronomy, Michigan State University, East Lansing, Michigan 48824, USA}

\author{W. Nazarewicz\,\orcidlink{0000-0002-8084-7425}}\email{Email: witek@frib.msu.edu}
\affiliation{Facility for Rare Isotope Beams, Michigan State University, East Lansing, Michigan 48824, USA}
\affiliation{Department of Physics and Astronomy, Michigan State University, East Lansing, Michigan 48824, USA}

\begin{abstract}
Recent measurements of the $E2$ transition rate from the ground state to the first 2$^+$ excited state of the proton dripline nucleus $^{36}$Ca show an unusual pattern when compared to its isotopic neighbor $^{38}$Ca: despite having a higher $E_x(2_1^+)$ excitation energy, the $B(E2; 0^+_1\rightarrow 2^+_1)$ rate in $^{36}$Ca is larger. 
The question that naturally arises is to what extent this observation can be attributed to the unbound character of the $2^+_1$ state. To understand the influence of the continuum space on the low-energy properties of $^{36}$Ca, we carried out Gamow shell model calculations that can account for the continuum coupling effects associated with the occupation of unbound $fp$ shells.
We found that in the threshold $2^+$ state, $^{36}$Ca is spatially diffused,
which impacts the observed $B(E2)$ trend. 
\end{abstract}

\date{\today}

\maketitle
\end{CJK*}

{\it Introduction.}--
The neutron-deficient Ca isotopes have attracted attention in recent years. Progress in experimentation at rare-isotope facilities allowed for the first Penning trap mass measurement of \nuc{36}{Ca}~\cite{Surbrook2021}, the precision charge-radius measurements of \nuc{36,37,38}{Ca} \cite{Miller2019}, one- and two-neutron transfer-reaction studies induced by \nuc{37}{Ca} \cite{Lalanne2021,Lalanne2022,Lalanne2023}, invariant mass spectroscopy providing the ground-state masses of the previously unknown \nuc{37,38}{Sc} and \nuc{34}{K} \cite{Dronchi2024}, a two-neutron knockout measurement leading to \nuc{36}{Ca} \cite{Beck2023}, as well as the determination of the $B(E2; 0^+_1 \rightarrow 2^+_1)$ transition strengths for \nuc{36,38}{Ca} \cite{Dronchi2023}. A picture emerged in the literature where shell-model (SM) calculations that allow for significant proton excitations from the $sd$ shell across the $Z=20$ gap into the $fp$ orbitals \cite{Caurier2001} describe the two-neutron knockout cross sections of \cite{Beck2023} and the surprisingly large $B(E2; 0^+_1\rightarrow 2^+_1) \equiv B(E2 \uparrow)$ transition strength reported for \nuc{36}{Ca} \cite{Dronchi2023}, perhaps corroborated independently by a study of the separation energies indicating a reduced $Z=20$ shell gap \cite{Dronchi2024}. However, the structure of \nuc{35,36}{Ca} must be driven by an interplay between shell evolution along $Z=20$ and the particle continuum, given their proximity to the proton dripline. Indeed, \nuc{35}{Ca} is the last proton-bound Ca isotope and the first $2^+$ state of \nuc{36}{Ca}, albeit predominantly decaying by $\gamma$-ray emission \cite{Lalanne2021,Dronchi2023}, is already a proton resonance above the particle threshold. Here, we will evaluate the impact of the particle continuum on the value of $B(E2 \uparrow)$ in \nuc{36}{Ca} and demonstrate that continuum effects indeed increase the transition strength.

The $2^+_1$ state of $^{36}$Ca has an excitation energy of 3.046(3) MeV \cite{Beck2023}; it lies slightly above the one- and two-proton thresholds at $S_p = 2.600(6)$ MeV~\cite{Surbrook2021,AME2021} and $S_{2p} = 2.650(40)$ MeV~\cite{AME2021}, respectively. Hence, continuum effects can influence this state and its decay. Indeed, it is expected that
the continuum-driven collectivization of near-threshold states may influence their electromagnetic decays \cite{Ploszajczak2020,Okolowicz2020b,Corbari2023,Bottoni2024}. 
The effects of the proton  continuum on structural properties of  $^{36}$Ca 
have been discussed in the context of the mirror energy difference between the $^{36}$Ca-$^{36}$S pair \cite{ValienteDobon2018,Okolowicz2008} and the charge radius of $^{36}$Ca \cite{Miller2019,Reinhard2022}.

To shed light on the experimental puzzle, we employ the Gamow shell model (GSM) \cite{Michel2002,Michel2021}. The GSM framework is the configuration mixing approach with the continuum incorporated in a self-consistent way \cite{Michel2009,Michel2023}; it has been widely used to study decay properties of threshold states~\cite{Michel2021,Dong2022}. We use this tool to investigate the abnormal electromagnetic transitions in $^{36,38}$Ca.

{\it Method.} --- Our GSM framework employs  valence-space effective operators~\cite{Xu2023} that are used to renormalize electromagnetic transition operators within the Berggren basis~\cite{Berggren1968,Michel2002}. Starting from chiral two-nucleon forces (2NF) and three-nucleon forces (3NF), the intrinsic Hamiltonian for an $A$-nucleon system can be written as:
\begin{equation}
    H = \sum_{i<j}^{A}\frac{(\vec{p}_{i} - \vec{p}_{j})^2}{2mA} + \sum_{i<j}^{A} v_{ij}^\text{NN} +\sum_{i<j<k}^{A}v_{ijk}^\text{3N},
    \label{eq1}
\end{equation}
where $\vec{p}_i$ represents the momentum of the $i$-th nucleon in the laboratory frame, and $m$ denotes the mass of the nucleon.

To account for the continuum effects, the Berggren basis, generated via the spherical Gamow-Hartree-Fock (GHF) approximation \cite{Hagen2006,Zhang2023}, is employed.  Utilizing the Berggren basis in conjunction with many-body perturbation theory (MBPT) \cite{Kuo1971,Coraggio2020}, we consistently construct both the complex valence-space effective Hamiltonian and the corresponding effective operators
closely following Ref.\,\cite{Xu2023}.

{\it Hamiltonian and model space.} --- To adequately describe the threshold effects, we adopted an optimized chiral force based on the  interaction EM1.8/2.0~\cite{Machleidt2011,Hebeler2011}. The nucleus $^{28}$Si serves as the reference state for GHF calculations and the GSM core. The core and valence spaces are those of the ZBM2 interaction~\cite{Zuker2015}. Before renormalization, we use the harmonic oscillator (HO) basis to evaluate the interaction matrix elements, with $\hbar\omega=16$ MeV, 13 major shells ($e=2n+l\leq e_{\rm max}=12$), and $e_{3\text{max}}=e_1+e_2+e_3\leq12$ for 3NF. GHF calculations yield bound states for the $\pi 1s_{1/2}$ and $\pi 0d_{3/2}$ orbits within the discrete, real-energy HF basis. The $\pi 0f_{7/2}$ orbit is resonant as it lies above the Coulomb barrier~\cite{Miller2019}. Also, the $\pi 1p_{3/2}$ orbit exhibits resonant behavior in the GHF calculation; both shells are incorporated via the complex-momentum GHF basis. The complex-momentum contours for $\pi f_{7/2}$ and $\pi p_{3/2}$ partial waves are defined as $k=0 \rightarrow 0.65-0.20i \rightarrow 1.30 \rightarrow 4$ fm$^{-1}$ and $k=0 \rightarrow 0.55-0.30i \rightarrow 1.20 \rightarrow 4$ fm$^{-1}$, respectively. Each partial wave is discretized with 35 scattering states. The valence space for the GSM calculations includes the $\pi1s_{1/2}$ and $\pi0d_{3/2}$ shells, resonant $\pi1p_{3/2}$ and $\pi0f_{7/2}$ orbitals with the corresponding scattering continua, as well as the $\nu1s_{1/2}$, $\nu0d_{3/2}$, $\nu1p_{3/2}$, and $\nu0f_{7/2}$ bound shells. We have checked that the inclusion of the $\pi s_{1/2}$ scattering continuum does not impact our results.
The MBPT calculation yielded energies $-5.054$\,MeV for the $\pi 1s_{1/2}$ bound state and $-4.567$\,MeV for the $\pi 0d_{3/2}$ bound state, while the wave numbers were $(0.427, -0.047)~\mathrm{fm}^{-1}$ for the $\pi 1p_{3/2}$ resonant state and $(0.513, -0.001)~\mathrm{fm}^{-1}$ for the $\pi 0f_{7/2}$ resonant state. We allow at most two valence particles in the scattering continuum, which guarantees converged results. To investigate the influence of the continuum, we employ the standard shell model with the same interaction used in GSM but solved in the localized real-space HF basis. In the following, we refer to this variant as SM.

{\it Interaction optimization.} --- To effectively capture the low-lying states and proton thresholds in $^{36,38}$Ca -- key to understanding the continuum's impact on electromagnetic properties -- the interaction's optimization is crucial. We first calibrated the low-energy constants (LECs) of the chiral EFT interaction with the conventional SM calculation. Subsequently, we fine-tuned these parameters within the GSM.

The calibration was performed using the POUNDerS method, which is a derivative-free optimization framework~\cite{petsc}. This involves iteratively adjusting the LECs to generate the full-space nuclear force, followed by applying MBPT to derive the valence space interaction. Considering the computationally intensive and time-consuming nature of the entire workflow, we utilized an emulator to accelerate the process by efficiently solving the Schr\"{o}dinger equation for various sets of LECs.

Our objective was to efficiently determine the energy $E_{\odot}\equiv E(\vec{\alpha}_{\odot})$ and wave unction $\psi_{\odot}\equiv \psi(\vec{\alpha}_{\odot})$ of the Hamiltonian $\hat{H}(\vec{\alpha}_{\odot})$, given a set of target LECs $\vec{\alpha}_{\odot}$. 
To this end, we employed the Eigenvector Continuation (EC) method, a powerful reduced basis technique for constructing emulators~\cite{Frame2018}.
The core principle of EC involves using a reduced basis of $N$ wave functions $\ket{\psi_i}$, obtained from high-fidelity SM calculations at various sets of LECs ${ \vec{\alpha}_i}$. Consequently, for any $\vec{\alpha}_{\odot}$, the subspace Hamiltonian matrix $\hat{\mathbb{H}}$ is expressed as:
\begin{equation}
\langle \psi_i | \hat{H}(\vec{\alpha}_{\odot}) | \psi_j \rangle = \langle \psi_i | T + V_0 | \psi_j \rangle + \sum_{k=1}^{N_{\rm LECs}} \alpha_{\odot, k} \langle \psi_i | V_k | \psi_j \rangle
\end{equation}
where the terms $\langle \psi_i | T + V_0 | \psi_j \rangle$ and $\langle \psi_i | V_k | \psi_j \rangle$ are invariant with respect to $\alpha_k$ in a linear system. Thus, they can be pre-computed at the outset, making the computational process more efficient. 

\begin{figure}[htb]
\includegraphics[width=1\columnwidth]{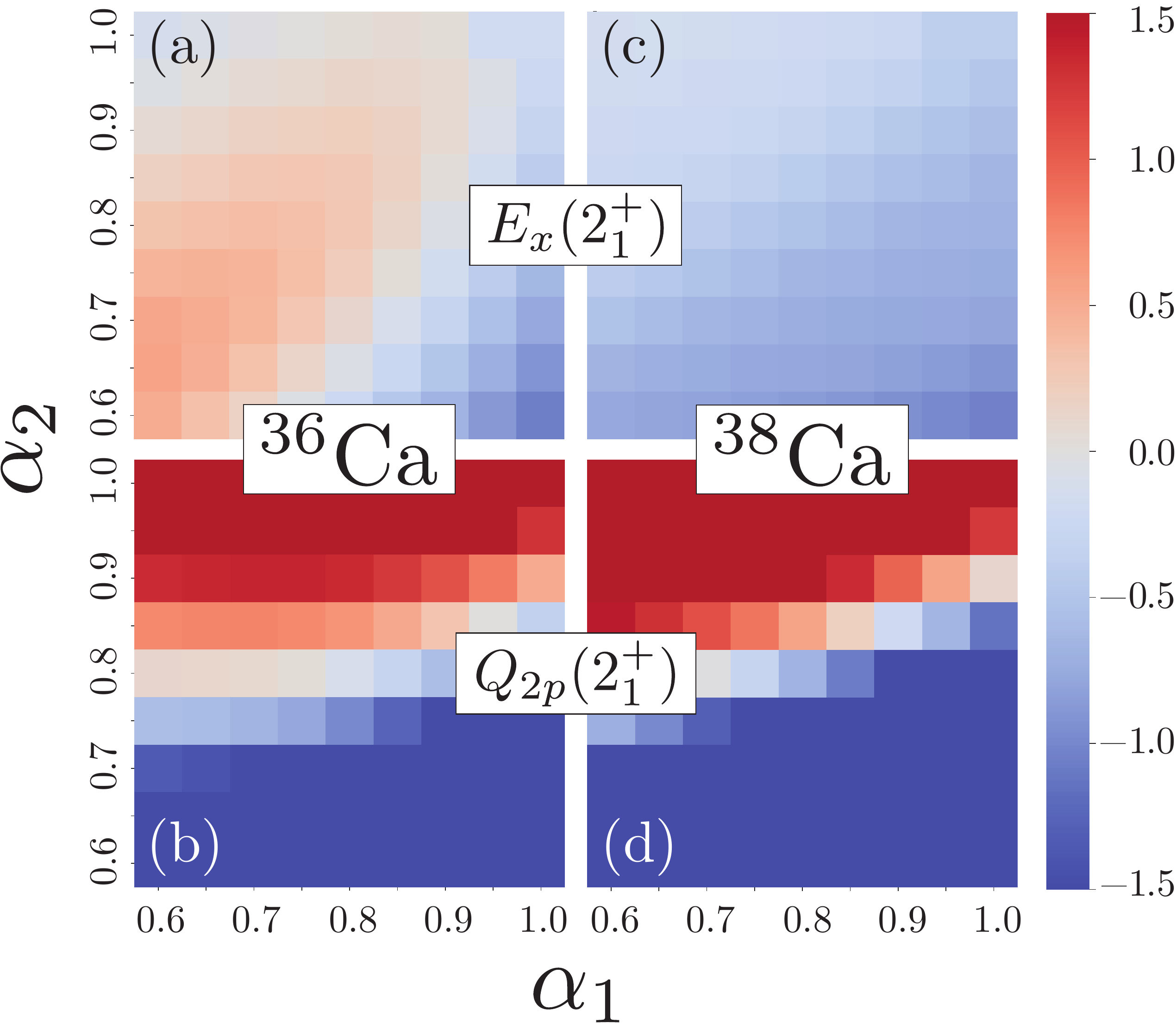}
\caption{The differences between the calculated SM values 
 of the $2_1^+$ excitation energy $E_x(2_1^+)$ (top) and the two-proton decay energy $Q_{2p}$ (bottom)
and experiment
for $^{36}$Ca (left) and $^{38}$Ca (right).
The parameters $\alpha_1$ and $\alpha_2$ are scaling factors applied to the $\tilde{C}_{^1 S_{0}}^{np}$ and $C_{^3 S_{1}- ^3 D_{1}}$ LECs. The experimental data are taken from Refs.~\cite{Beck2023,Beck2024,ENSDF}.}\label{SM_HF}
\end{figure}

In this study, we adjusted two LECs: the proton-neutron interaction strength $\tilde{C}_{^1 S_{0}}^{np}$ and the tensor force component $C_{^3 S_{1}- ^3 D_{1}}$, originally set at $-0.147167$ (in units 10$^4$\,GeV$^{-2}$) and $0.826$ (in units 10$^4$\,GeV$^{-4}$), respectively~\cite{Machleidt2011}. We trained emulators using 5 to 7 sample points of $\mathbf{p} = \{\tilde{C}_{^1 S_{0}}^{np}$, $C_{^3 S_{1}- ^3 D_{1}}\}$, each scaled by multiplicative factors $\alpha_1$ and $\alpha_2$ within the range of 0.6 to 1.0. The emulators aimed to reproduce the high-fidelity SM results shown in Fig.\,\ref{SM_HF}. These  heat maps show the deviations of the theoretical predictions from experimental values of the $2_1^+$ excitation energy $E_x(2_1^+)$ and the two-proton decay energy $Q_{2p}$. 

With the emulator, we use two scaling factors to adjust the values of $\tilde{C}_{^1 S_{0}}^{np}$ and $C_{^3 S_{1}- ^3 D_{1}}$, aiming to the accurate representation of $N_d=10$ observables $\mathcal{O}_i$ ($i = 1, \dots, 10$), which encompass the spectra and proton separation energies of $^{36,38}$Ca. The optimization procedure involved minimizing the penalty function:
\begin{equation}
\chi^2(\mathbf{p}) = \sum_{i=1}^{N_d} \left( \frac{\mathcal{O}_i(\mathbf{p}) - \mathcal{O}_i^{\text{exp}}}{\delta \mathcal{O}_i} \right)^2,
\end{equation}
where $\mathcal{O}_i(\mathbf{p})$ are computed observables and $\mathcal{O}_i^{\text{exp}}$ are experimental data used to constrain the model. For simplicity, the adopted errors $\delta \mathcal{O}_i$ were kept constant in this study.

The POUNDerS~\cite{petsc} is a derivative-free optimization approach that requires the specification of initial parameter values and an initial trust region. Utilizing the emulator based on the EC, we systematically sampled 1600 starting points, uniformly distributing the scaling factors $\alpha_1$ and $\alpha_2$ between 0.6 and 1.0 with a step size of 0.01. The initial trust region was set to a default value to promote a balanced exploration of the parameter space.

\begin{figure}[htb]
\includegraphics[width=1\columnwidth]{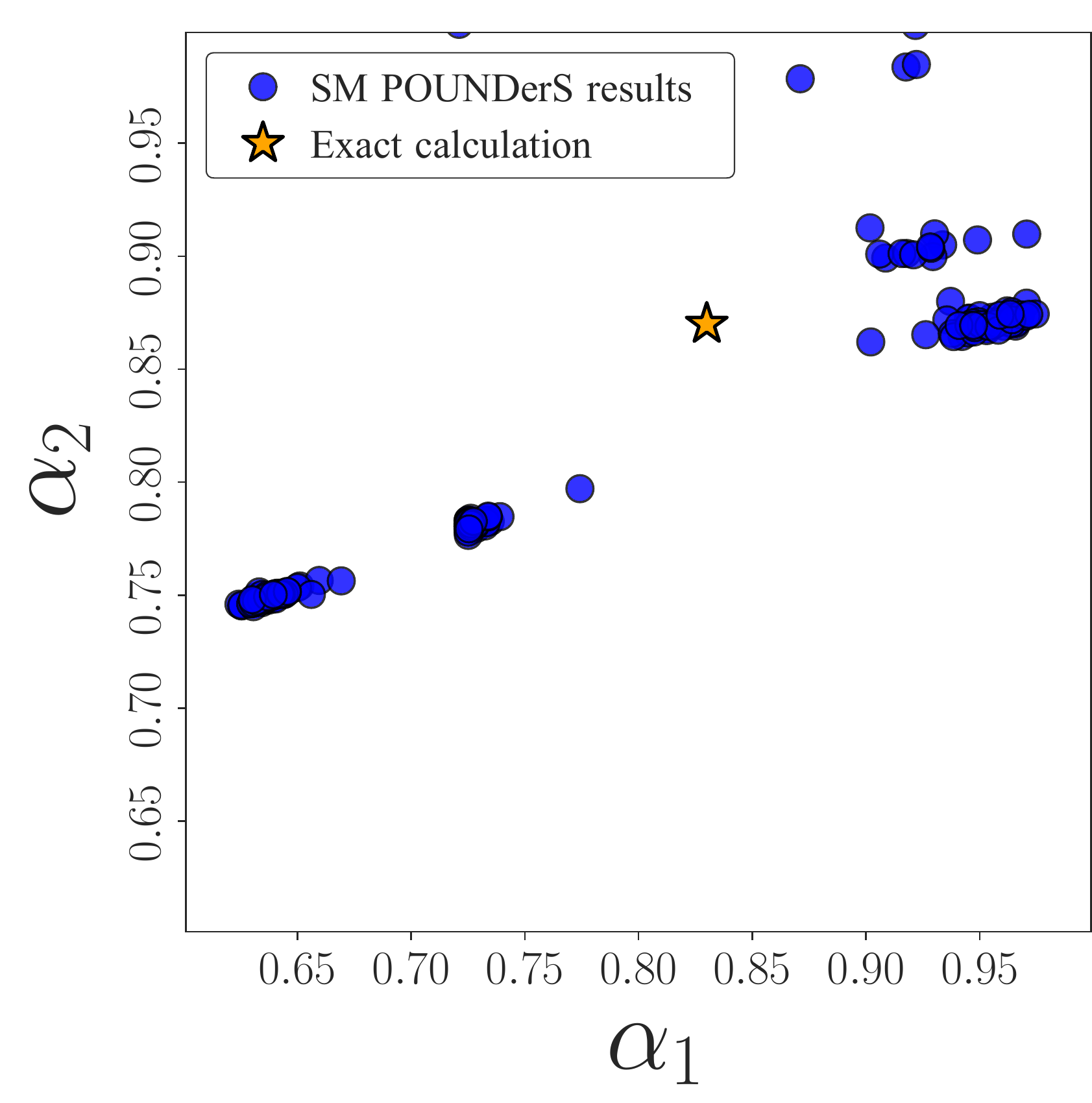}
\caption{The results from the POUNDerS optimization, displayed in the same coordinates as in Fig.\,\ref{SM_HF}. The blue points represent the outcomes from various starting points using the SM POUNDerS approach, while the  star indicates the values selected for the primary analysis.}\label{pounders}
\end{figure}

The optimization results are shown  in Fig.\,\ref{pounders}. Here, the blue dots represent the outcomes from POUNDerS at different starting points. These points closely cluster along a narrow band, aligning well with the previous high-fidelity SM calculations depicted in Fig.\,\ref{SM_HF}. 

We next conducted high-fidelity GSM calculations considering the continuum effects, focusing on LEC sets within this identified band. As a result, the optimal values of $\alpha_1$ and $\alpha_2$ were determined to be 0.83 and 0.87, respectively, as indicated by the  star in Fig.\,\ref{pounders}. These values provide a good description of the spectra for both SM and GSM calculations. The star is positioned slightly above the blue band, because the inclusion of the proton continuum typically lowers the energy of the $2^+_1$ state.

The  resulting  two chiral LECs are: $\tilde{C}_{^1 S_{0}}^{np} = -0.122$ (in units 10$^4$\,GeV$^{-2}$) and $C_{^3 S_{1}- ^3 D_{1}} = 0.719$ (in units 10$^4$\,GeV$^{-4}$). The  GSM and SM calculations presented in this work are based on the same optimized chiral interaction. The only difference between the two lies in the basis used: the GSM utilizes the GHF  basis, which includes the continuum space, whereas the SM employs a localized real-space Hartree-Fock basis. In both calculations, the effective Hamiltonian and effective operators derived from MBPT are used, without explicitly introducing any effective charges.

\begin{figure}[htb]
\includegraphics[width=1\columnwidth]{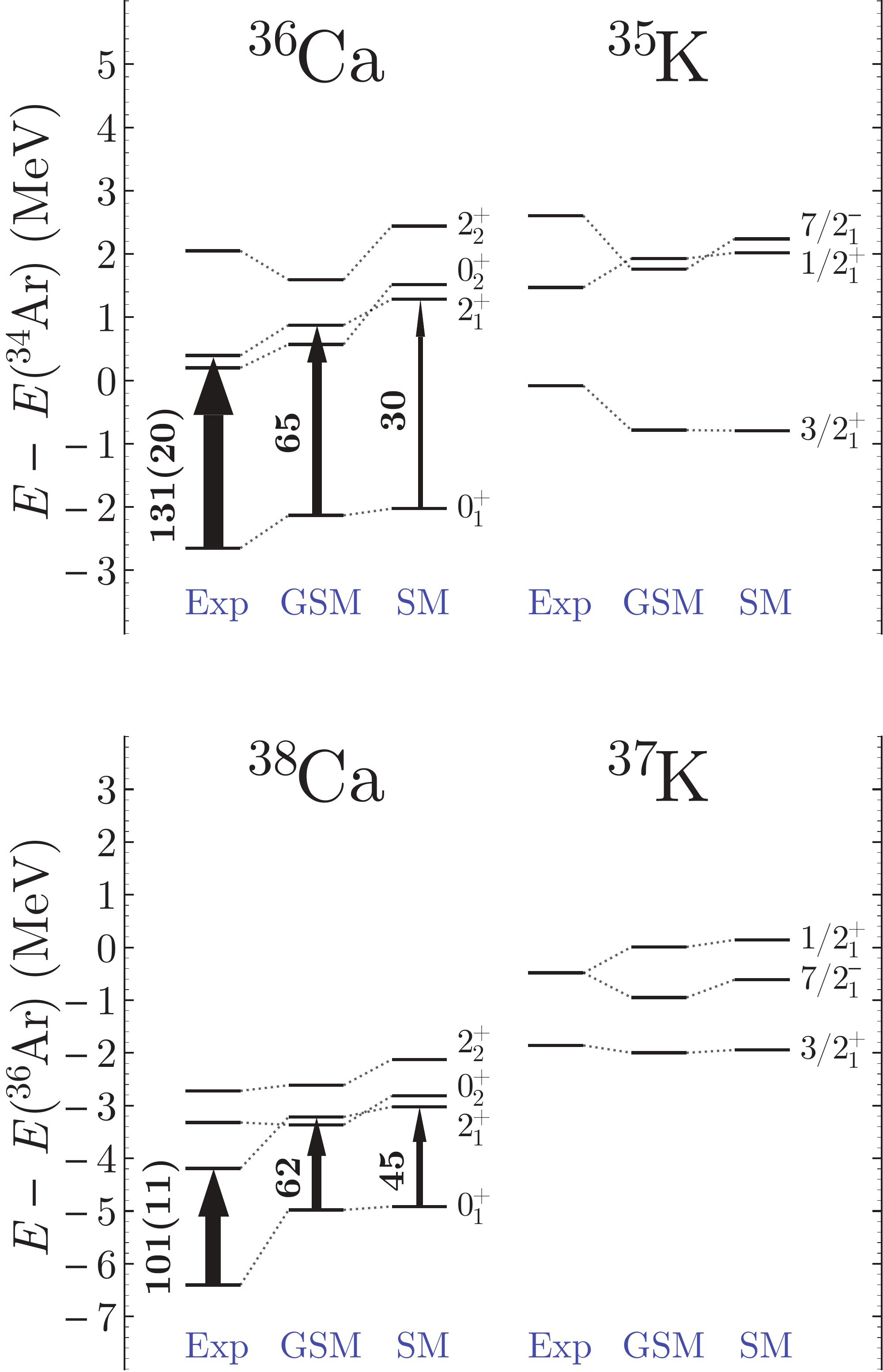}
\caption{Calculated spectra for $^{36,38}$Ca and $^{35,37}$K obtained using GSM and SM methods. 
The $B(E2)$ values (in $e^2$fm$^4$) for the  $0^+_1 \rightarrow 2^+_1$ transitions are marked. Experimental data are taken from Refs.\,\cite{Dronchi2023,Beck2023,Beck2024}.
}\label{Spectra}
\end{figure}
{\it Results.} --- The calculated spectra of $^{36,38}$Ca and $^{35,37}$K reasonably agree with experimental data, as illustrated in Fig.\,\ref{Spectra}. However, the $B(E2)$ values obtained in both SM and GSM approaches are smaller than in the experiment.
In the conventional SM analysis with phenomenological interactions, the $B(E2)$ values are usually calculated using effective charges. 
In our MBPT framework, the orbital-dependent effective charges can be obtained by renormalizing the electromagnetic transition operator self-consistently~\cite{Xu2024}, resulting in $e_p \approx 1.05, e_n \approx 0.25$. These values are significantly lower compared to those from calculations using the phenomenological ZBM2 interaction with  $e_p = 1.36$ and $e_n = 0.45$ \cite{Dronchi2023}.
The primary reason of this discrepancy is the  truncation employed in the MBPT framework. Indeed, as discussed in Refs.~\cite{Stroberg2022,Sun2025}, the excitations out of the SM model space are crucial for explaining the missing $B(E2)$ strength.

\begin{figure}[htb]
\includegraphics[width=1\columnwidth]{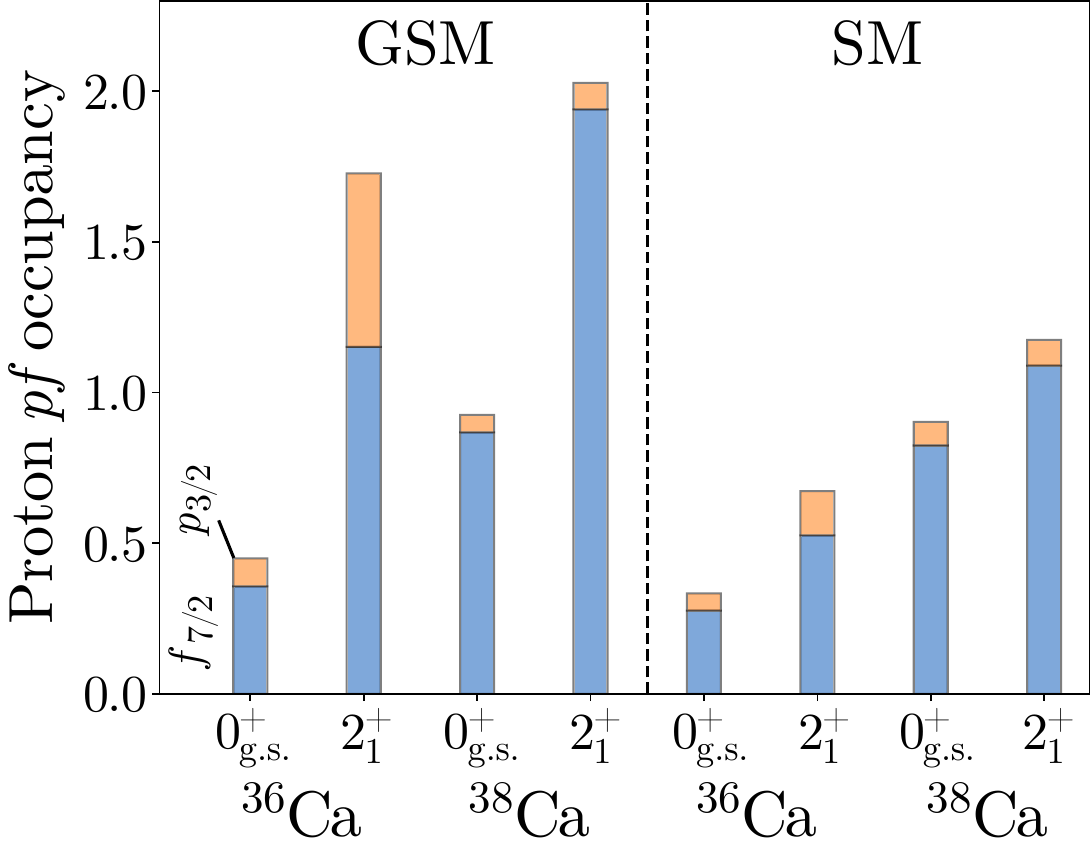}
\caption{Calculated proton $pf$-orbital occupancies of the $0_{\rm g.s.}^+$ and $2_1^+$ states in \nuc{36}{Ca} and \nuc{38}{Ca}, respectively, obtained using GSM and SM. In the GSM calculations, the occupancies of the unbound $f_{7/2}$ and $p_{3/2}$ orbitals are primarily determined by contributions from the resonant poles, with additional contributions from the scattering continuum.
}\label{occu}
\end{figure}

Nevertheless, the GSM $B(E2)$  values are significantly larger relative to the SM results. This enhancement is attributed to (i) the inclusion of the continuum in the GSM, which reduces the $Z=20$  gap, leading to increased cross-shell proton $sd \rightarrow pf$ excitations, and (ii) the extended character of continuum wave functions in the GSM. The effect (i)  significantly modifies the GSM configuration of the $2_1^+$ state in $^{36,38}$Ca as compared to the SM picture. As shown in Fig.\,\ref{occu}, while both models predict minor $pf$ occupancies in the $0^+_1$ ground states of $^{36,38}$Ca, the GSM framework exhibits a notable increase in $pf$ occupancies in the $2_1^+$ states. In particular, the $2^+_1$ state of $^{36}$Ca shows a pronounced enhancement of the $p$-wave component, whereas the $f$-wave component becomes dominant in $^{38}$Ca. The total occupation numbers of the $f_{7/2}$ and $p_{3/2}$ partial waves in the GSM approach are between 1.7 and 2 for the $2_1^+$ states of $^{36,38}$Ca. Given that these partial waves correspond to the proton continuum, this highlights the significance of the contributions from two particles in the continuum, including resonant poles.

This cross-shell effect also influences the $0_2^+$ states of \nuc{36,38}{Ca}. As highlighted in Ref.\,\cite{Heyde2011}, the $0_2^+$ states in Ca isotopes and the mirror nucleus \nuc{36}{S} feature significant intruder and cluster components. These components can be partially incorporated in GSM calculations through the renormalization of the large-space Hamiltonian. In the GSM results, the $0_2^+$ state is characterized by a two-particle-two-hole proton intruder configuration, with proton occupancies of 0.69 in the $p_{3/2}$ orbital and 1.19 in the $f_{7/2}$ orbital. This is consistent with previous studies \cite{Lalanne2022,ValienteDobon2018} which used a SM with an enhanced N$\hbar \omega$ component in the SDPF-U-mix interaction to simulate the cross-shell effect.

\begin{figure}[htb]
\includegraphics[width=1\columnwidth]{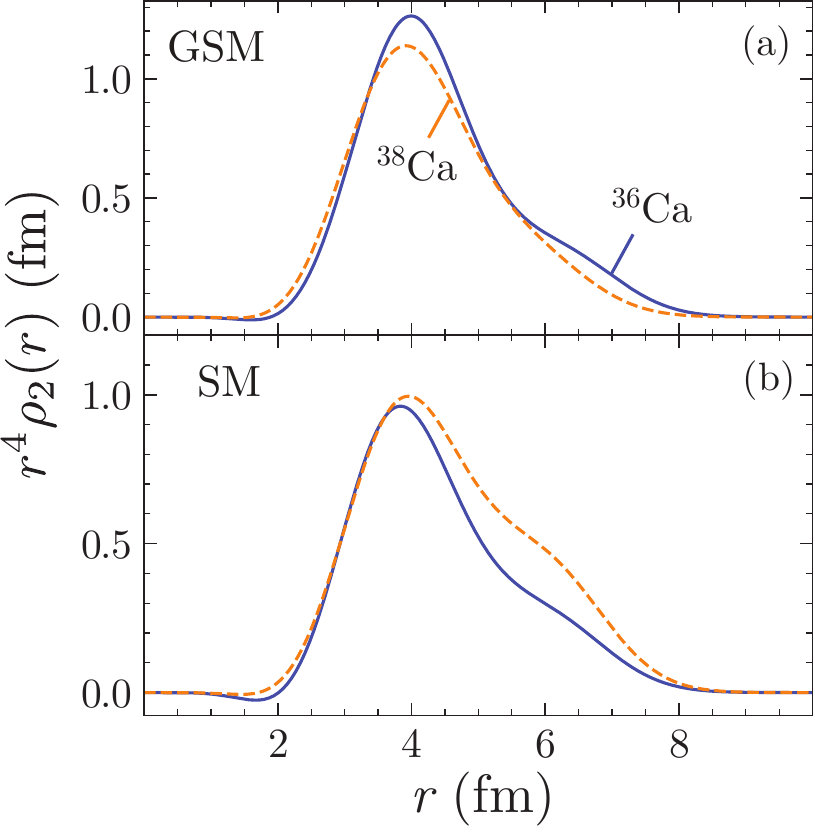}
\caption{Predicted $E2$ transition densities from the ground state to the 2$^+_1$ state for $^{36}$Ca (solid lines) and $^{38}$Ca (dashed lines). Results were obtained with (a) GSM and (b) SM.}\label{Transition_density}
\end{figure}

As shown in Fig.~\ref{Spectra}, the proton continuum impacts the $B(E2)$ values of $^{36}$Ca and $^{38}$Ca in a different
way. Specifically, in $^{36}$Ca, the inclusion of the continuum causes the $E2$ transition amplitudes for $p$ and $f$ waves to increase by over 50\%.  In $^{38}$Ca, only the transition amplitude for the $f$ wave  increases by 37\%. This is due to the fact that the unbound $p$ wave, having a lower centrifugal barrier, leads to a more diffuse spatial distribution.  To illustrate this, we compute
the $E2$ transition density $\rho_2$ that defines  the $B(E2)$ rate \cite{Heisenberg1981}:
\begin{equation}
    B(E2; 0^+ \rightarrow 2^+) = 5 e^2 \left[\int_{0}^{\infty}\rho_{2}(r) r^{4}   dr\right]^2.
\end{equation}

Figure\,\ref{Transition_density}   shows $r^4 \rho_2(r)$, 
obtained in both models for $^{36}$Ca and $^{38}$Ca. 
According to our calculations, the transition densities for \nuc{36}{Ca} and \nuc{38}{Ca} are remarkably similar within the nuclear interior. However,  the transition density for $^{36}$Ca exhibits a visible enhancement at larger distances compared to $^{38}$Ca. An opposite situation is predicted in the SM calculation: it is the transition density for \nuc{38}{Ca} that is larger in the surface region.
The halo pattern  shown in Fig.~\ref{Transition_density} is characteristic of weakly-bound or unbound systems, in which the nuclear wave function tends to extend beyond the nuclear surface.

% \WN{Possible discussion of the transition density here.}

{\it Summary.} ---  Using configuration interaction based GSM that can account for continuum effects, we study low-energy properties of $^{36,38}$Ca. We demonstrate that the  abnormally large $B(E2; 0^+ \rightarrow 2^+)$ value in \nuc{36}{Ca} can be partly explained in terms of the low-lying proton continuum effect. In particular, 
our  analysis reveals that the proton $sd\rightarrow fp$ excitations  strongly influence the 
$B(E2; 0^+ \rightarrow 2^+)$ transition. 

In the weakly-bound $^{36}$Ca, the inclusion of the continuum space significantly enhances cross-shell excitations and proton occupancies in the unbound $p$ and $f$ partial waves. While the current level of numerical precision of the GSM makes it challenging to directly extract a proton decay rate, we do find that the continuum effect -- particularly the presence of the significant $p$-wave component in the unbound $2^+$ state -- leads to an increased $E2$ transition density around and beyond the nuclear surface. This underscores the significant impact of the continuum on electromagnetic transitions, particularly in 
weakly-bound dripline systems. We note that a similar conclusion regarding the $B(E2)$ strength in $^{36}$Ca was obtained in 
the shell model embedded in the continuum (SMEC) framework of Ref.~\cite{Okolowicz2024}, where, however, only one proton was allowed to scatter into the non-resonant continuum. 

We wish to emphasize that other effects, such as the excitations out of the employed $sd-pf$ SM model space, that result in increasing the low-energy $B(E2)$ strength through the coupling to the giant quadrupole resonance, are essential for reproducing the experimental $E2$ transition rates, see, e.g.,  discussion in Refs.~\cite{Stroberg2022,Sun2025,Dytrych2020,McCoy2020,McCoy2025}.

Experimentally, the evolution of the $Z=20$ shell closure towards the proton dripline
can be investigated through a variety of observables, such as trends in proton separation energies
\cite{Dronchi2024,Buskirk2024}. Their differences highlight the proton-magic character of \nuc{40}{Ca} and hint at increased stability at $Z=20$ for the $N=19$ isotones, which appears to vanish for $N=18$~\cite{Dronchi2024}. Mass measurements of the proton-unbound \nuc{42}{V} ($N=19$) and \nuc{41}{V} ($N=18$) will enable tracing the weakening of the $Z=20$ shell closure towards the proton dripline.
Also, the measurement of spectroscopic factors for proton removal from $^{36,38}$Ca to potentially proton-unbound, $pf$-shell final states of $^{35,37}$K~\cite{Beck2024} is called for as it can provide more detailed information on the ground-state configurations of $^{36,38}$Ca.

{\it Acknowledgments.} --- This material is based upon work supported by the National Key Research and Development Program (MOST 2023YFA1606404 and MOST 2022YFA1602303); the National Natural Science Foundation of China under Contract No.\,12347106, No.\,12147101, and No.12447122;  the China Postdoctoral Science Foundation under Contract No.\,2024M760489 
and by the U.S. Department of Energy under Award Numbers DE-SC0013365 and DE-SC0023633 (Office of Science, Office of Nuclear Physics) and DE-SC0023175 (Office of Science, NUCLEI SciDAC-5 collaboration).

\bibliography{references}
\end{document}